\documentclass[twocolumn,showpacs,preprintnumbers,amsmath,amssymb]{revtex4}
\usepackage{amsmath,amsfonts,diagbox,latexsym,amssymb,graphics,epsfig,subfigure,color,makeidx}
\usepackage{xcolor}
\usepackage[utf8]{inputenc}
\usepackage{graphicx,hyperref}
\usepackage{bm}
\usepackage{color}
\usepackage{multirow}

\begin{document}
	\title{Evolution and disruption of circular orbits during dynamic black hole scalarization}
	\author{Yu-Peng Zhang\footnote{zhangyupeng@lzu.edu.cn},	
		    Shao-Wen Wei\footnote{weishw@lzu.edu.cn},
		    Yu-Xiao Liu\footnote{liuyx@lzu.edu.cn, corresponding author}
	}
	\affiliation{
        Lanzhou Center for Theoretical Physics, Key Laboratory of Theoretical Physics of Gansu Province, Key Laboratory of Quantum Theory and Applications of MoE, Gansu Provincial Research Center for Basic Disciplines of Quantum Physics,\\
Institute of Theoretical Physics \& Research Center of Gravitation, School of Physical Science and Technology, Lanzhou University, Lanzhou 730000, China
	}
	
	\begin{abstract}
		
Dynamical black hole scalarization describes the process by which a ``no-hair" black hole transitions to a scalarized hairy state. This transition significantly alters the spacetime geometry, including the structure of stable circular orbits of the test particles, which are critical for astrophysical observations and tests of gravity. In this work, we employ numerical relativity simulations to model the time-dependent evolution of a vacuum black hole undergoing scalarization. By evolving timelike geodesic equations within the dynamically changing spacetime background, we investigate how the scalarization process impacts initially stable circular orbits. Our results reveal that the growth of the scalar hair destroys the circular nature of these orbits, forcing them to either become eccentric or plunge into the black hole. This orbital destabilization arises from the rapid change of spacetime curvature induced by the scalar field. Our findings provide novel insights into the interplay between scalar fields and black hole dynamics, with significant implications for gravitational-wave signatures, accretion disk stability, and observational tests of extended theories of gravity.
		
	\end{abstract}
	\maketitle
	\section{Introduction}
	
Black holes, once confined to the realm of theory, have moved to the center stage of modern astrophysics, supported by robust observational evidence. The Event Horizon Telescope has successfully imaged the shadows of M87* \cite{eth2019} and Sagittarius A* \cite{eth2022}, while gravitational-wave detectors such as LIGO/Virgo/KAGRA have cataloged numerous binary black hole mergers \cite{Abbott2016a,LIGOScientific:2018jsj,LIGOScientific:2020kqk,LIGOScientific:2021psn}. These breakthroughs confirm the physical existence of black holes and validate general relativity in the strong-field regime. They are consistent with the no-hair theorem, a cornerstone of general relativity, which asserts that isolated, stationary black holes are uniquely characterized by only three parameters: mass ($M$), angular momentum ($J$), and electric charge ($Q$) \cite{Chrusciel:2012jk}. In vacuum spacetimes, this theorem implies that non-rotating black holes are described by the Schwarzschild metric, while rotating ones are governed by the Kerr solution, precluding the existence of ``additional hair" such as scalar fields.

Early investigations into scalar fields reinforced the no-hair conjecture. Specifically, Chase was the first yo demonstrate that a static, asymptotically flat black hole cannot support static, massless, minimally coupled scalar hair in general relativity \cite{Chase:1970omy}. Bekenstein and Hawking subsequently generalized this result, proving that even self-interacting scalar fields cannot form hair around stationary, asymptotically flat black holes \cite{Hawking:1972qk, Bekenstein:1971hc, Bekenstein:1972ky, Bekenstein:1995un}. These no-scalar-hair theorems established that Schwarzschild black holes are strictly vacuum solutions in standard general relativity. However, this conclusion no longer holds when the assumptions of the no-hair theorems are relaxed. Novel hairy solutions have been constructed by incorporating non-Abelian gauge fields \cite{Volkov:1989fi, Bizon:1990sr, Greene:1992fw, Luckock:1986tr, Droz:1991cx}, introducing non-minimal couplings between the scalar field and curvature or electromagnetic invariants  \cite{Kanti:1995vq, Gregory:1992kr, Yamazaki:1989hy, Campbell:1991kz, Garfinkle:1990qj,Herdeiro:2018wub,Guo:2021zed, Yao:2021zid, Zhang:2021ybj, Zhang:2021nnn,Zhang:2024wci,Sotiriou:2013qea,Sotiriou:2014pfa, Benkel:2016kcq, Doneva:2017bvd, Silva:2017uqg, Andreou:2019ikc, Hod:2019pmb, Peng:2019snv, Ripley:2019aqj, Hod:2019vut, Dima:2020yac,Tang:2020sjs, Liu:2020yqa, Guo:2020sdu, Herdeiro:2020wei, East:2021bqk, Zhang:2022kbf, Wang:2020ohb,Lin:2023npr}, or considering time-dependent, complex scalar fields in synchronized with the horizon of rotating or charged black holes \cite{Herdeiro:2014goa, Sanchis-Gual:2015lje}.

Spontaneous scalarization, originally introduced by Damour and Esposito-Far\`ese in the context of neutron stars \cite{Damour:1993hw}, describes a phase transition where a compact object develops a nontrivial scalar hair due to a tachyonic instability induced by non-minimal couplings. This mechanism has been successfully extended to black hole spacetimes, revealing a rich phenomenology \cite{Benkel:2016kcq,Andreou:2019ikc,Guo:2021zed,Zhang:2021ybj,Zhang:2021nnn,Fernandes:2019rez,Herdeiro:2018wub,East:2021bqk}. A key feature of black hole scalarization is its dynamical nature: a vacuum, no-hair black hole can continuously transition into a scalarized state when specific criteria (such as a critical coupling strength) are met. During this transition, the spacetime curvature evolves from a standard Schwarzschild or Kerr geometry to a hairy configuration. Crucially, this process is inherently non-equilibrium; the exponential growth of the scalar field exerts a significant backreaction on the metric.

The spacetime structure surrounding a black hole governs the motion of both matter and light. Two critical geometric features define this environment: the innermost stable circular orbit, which marks the boundary between stable accretion and the plunge into the horizon, and the photon sphere (or light rings), which consists of unstable null geodesics. These structures play a decisive role in astrophysical processes, regulating accretion disk dynamics, jet formation, and the gravitational-wave signals from inspiraling compact binaries. In standard general relativity, the ``no-hair" property ensures that these orbits are determined uniquely by the black hole's mass and spin. However, scalarization fundamentally challenges this universality. As the black hole transitions to a hairy state, the background geometry evolves dynamically, shifting the locations of the innermost
stable circular orbit and photon sphere. Consequently, initially stable circular orbits may lose their stability or develop eccentricity. Such dynamical modifications could leave distinct imprints on observables—including shifts in accretion disk thermal spectra, quasar variability, or phase deviations in gravitational waveforms—providing a potential window to distinguish general relativity from modified gravity theories.

In this work, we investigate the dynamical coupling between black hole scalarization and orbital motion using full numerical relativity simulations. We model the continuous transition of an initially vacuum ``no-hair" black hole into a scalarized state within a fully general relativistic framework, capturing the nonlinear, time-dependent evolution of both the spacetime metric and the scalar field. To probe the orbital dynamics, we integrate the geodesic equations using the $3+1$ decomposition formalism \cite{Vincent:2012kn,Bohn:2014xxa}, tracking test particles initially placed in stable circular orbits around the seed black hole. By quantifying the deformation of these orbits—specifically, the induced eccentricity and the plunge behavior caused by the dynamically changing curvature—we address a critical gap in understanding how the scalarization process alters fundamental orbital properties. Our results bridge the theoretical predictions of scalarized geometries with their potential astrophysical consequences, offering a robust framework to interpret future observations of black holes in modified theories of gravity.

This paper is organized as follows. In Sec.~\ref{sec:fundamentals}, we outline the framework of Einstein-Maxwell-dilaton gravity and present the corresponding equations of motion for both the gravitational and matter fields. Section~\ref{sec:results} details our numerical simulation results and provides a physical analysis of the orbital dynamics. Finally, we summarize our findings and discuss future prospects in Sec.~\ref{Conclusion}.
	
\section{Setup} \label{sec:fundamentals}

We consider an Einstein-Maxwell-dilaton model described by the following action \cite{Garfinkle:1990qj,Zhang:2021ybj}
	\begin{equation}
	S = \int d^4x \sqrt{|g|}
	\bigg(\frac{R-F(\phi)\mathcal{I}}{16\pi}-\frac{1}{2}\partial_\mu\phi \partial^\mu\phi-V(\phi)\bigg).
	\label{action}
	\end{equation}
Here, $g$ is the determinant of the metric, $R$ is the Ricci scalar, and the electromagnetic invariant is defined as
\begin{equation}
\mathcal{I} \equiv F_{\alpha\beta}F^{\alpha\beta},
\end{equation}
where the electromagnetic tensor is
\begin{equation}
F_{\alpha\beta} = \nabla_\alpha A_\beta - \nabla_\beta A_\alpha.
\end{equation}
The coupling function $F(\phi)$ governs the interaction between the scalar and electromagnetic fields and determines the scalarization properties of the system. Specifically, if the derivative vanishes at the background value
\begin{equation}
dF/d\phi|_{\phi_0}=0,
\end{equation}
the theory admits the un-scalarized standard Kerr-Newman black hole as a valid solution \cite{Herdeiro:2018wub,Fernandes:2019rez}.

In this paper, we focus on an exponential coupling function
\begin{equation}
F(\phi)=e^{\eta \phi},
\label{fofphi}
\end{equation}
where $\eta$ is the coupling parameter. For this specific choice, the derivative $dF/d\phi$ never vanishes.  Consequently, unlike models with quadratic coupling (where $dF/d\phi|_{\phi_0}=0$), our model does not admit a static, scalar-free no-hair black hole as an equilibrium solution. The non-vanishing coupling term acts as a constant source for the scalar field. Nevertheless, the standard charged black hole solution (Kerr-Newman or Reissner-Nordstr\"{o}m) serves as the initial configuration for our dynamical simulations, from which the scalarization process naturally evolves. Throughout this paper, we adopt geometric units where $G=c=\hbar=1$.
	
Varying the action \eqref{action} with respect to the metric $g_{\mu\nu}$, the electromagnetic potential $A_\mu$, and the scalar field $\phi$ yields the following equations of motion
\begin{eqnarray}
	R_{\mu\nu}-\frac{1}{2}g_{\mu\nu}R &=& 8\pi \left(T_{\mu\nu}^{(\phi)} + T_{\mu\nu}^{(\text{em})}\right),
	\label{einsteineq}\\
	\nabla_\mu\left(e^{\eta \phi}F^{\mu\nu}\right) &=& 0,
	\label{maxwelleq}\\
	\nabla_\mu\nabla^\mu\phi &=& \frac{\eta}{16\pi}\mathcal{I}e^{\eta \phi}+\frac{dV(\phi)}{d\phi}.
	\label{keleineq}
\end{eqnarray}
Here, $T^{(\phi)}_{\mu\nu}$ and $T_{\mu\nu}^{(\text{em})}$ denote the energy-momentum tensors for the scalar field and the electromagnetic field, respectively. They are defined as
\begin{equation}
	T^{(\phi)}_{\mu\nu} = \partial_\mu\phi\partial_\nu\phi - g_{\mu\nu}\left(\frac{1}{2}\partial^\alpha\phi\partial_\alpha\phi+V(\phi)\right),
\end{equation}
and
\begin{equation}
	T_{\mu\nu}^{(\text{em})} = \frac{1}{4\pi}e^{\eta \phi}\left(F_{\mu\alpha}F_\nu^{~\alpha}-\frac{1}{4}g_{\mu\nu}F^{\alpha\beta}F_{\alpha\beta}\right).
\end{equation}
We note that in the limit $\eta \to 0$, the non-minimal coupling vanishes, and the model reduces to the standard Einstein-Maxwell theory minimally coupled to a scalar field.

In this paper, we restrict our analysis to spherically symmetric configurations. To solve the coupled system of field equations \eqref{einsteineq}, \eqref{maxwelleq}, and \eqref{keleineq}, we employ a dedicated spherical numerical relativity code \cite{Zhang:2023qag} based on the Baumgarte–Shapiro–Shibata–Nakamura (BSSN) formalism adapted to spherical polar coordinates ~\cite{Alcubierre:2011pkc,Montero:2012yr}. This approach allows us to capture the full nonlinear dynamics of the scalarization process. Specifically, we adopt the following metric ansatz
	\begin{eqnarray}
	ds^2&=&(-\alpha^2 + \beta^r \beta_r) dt^2 + 2\beta_r dt dr \nonumber\\
	&&+e^{4\chi}\left(a\,dr^2+b\,r^2 \,d\Omega^2 \right)
	\label{metric}
	\end{eqnarray}
with 	
	\begin{equation}
	d\Omega^2=(d\theta^2+\sin^2\theta d\varphi^2),
	\end{equation}
where $\alpha$, $\beta^i=(\beta^r, 0, 0)$, and $\chi$ are the lapse function, shift vector, and conformal factor, respectively. All the metric functions and fields are dependent on $(r, t)$. We can use the metric \eqref{metric} to obtain a 3-metric of the spacelike hypersurface $\Sigma_t$ as follows
	\begin{equation}
	\gamma_{ij}=e^{4\chi}\text{diag}\left(a, b\,r^2, b\,r^2\sin^2\theta\right).
	\end{equation}
For the sake of brevity, we do not explicitly list the full set of evolution equations for the BSSN metric functions, the scalar field, and the electromagnetic potential here. Detailed derivations and the specific forms of these equations can be found in Ref. \cite{Montero:2012yr}.
	
We initialize the system with a vanishing scalar field, such that the spacetime is initially described by the analytical Reissner-Nordstr\"om black hole with mass $M_0$ and charge $Q$. For the simulations presented in this work, we set the initial mass to $M_0=1$ and the charge to $Q=0.9$. Following the formulation in Ref.~\cite{Alcubierre:2009ij}, the conformal electric potential $\varphi$ and the conformal factor $e^{\chi}$ are given by
	\begin{equation}
	\varphi=\frac{Q}{r},
	\end{equation}
	\begin{equation}
	e^{2\chi}=\left[\left(1+\frac{M_0}{2r}\right)^2-\frac{Q^2}{4^2}\right].
	\end{equation}
The radial component of the physical electric field is given by
\begin{equation}
E^r=\frac{Q}{r^2 e^{6\chi}}.
\end{equation}
Consistent with the choice of isotropic coordinates for the initial data, the conformal metric functions are initially trivial, i.e., $a=b=1$.

\section{Dynamical evolution of timelike circular orbits} \label{sec:results}

Having established the numerical framework for the dynamical scalarization, we now turn to the analysis of orbital motion within this evolving background. In this section, we detail the procedure for integrating geodesics in dynamical spacetimes. Adopting the formalism developed in Ref.~\cite{Vincent:2012kn}, we present the geodesic equations decomposed within the $3+1$ framework. Consider a test particle $\mathcal{P}$ characterized by its 4-momentum $p^\mu$. Its trajectory defines a worldline—either null (for photons, where $p^\mu p_\mu=0$) or timelike (for massive particles, where $p^\mu p_\mu=-m^2$)—governed by the geodesic equation
\begin{eqnarray}
p^\mu\nabla_\mu p^\alpha=0.
\end{eqnarray}
To cast the geodesic equations into a form suitable for numerical integration, we perform an orthogonal decomposition of the 4-momentum $p^\mu$ with respect to the unit normal vector $n^\mu$ of the spatial hypersurfaces
\begin{eqnarray}
p^\mu = E(n^\mu + V^\mu),
\end{eqnarray}
where the spatial momentum vector $V^\mu$ satisfies the orthogonality condition $n_\mu V^\mu=0$. In this decomposition, $n^\mu$ represents the 4-velocity of the Eulerian observer $\mathcal{O}_E$, and the scalar $E$ corresponds to the energy of the particle $\mathcal{P}$ as measured by this observer. Substituting this decomposition into the normalization condition, we obtain
\begin{eqnarray}
E = - n^\mu p_\mu.
\end{eqnarray}
The vector $V^\mu$ represents the spatial velocity of the particle $\mathcal{P}$ as measured by the Eulerian observer $\mathcal{O}_E$. It is related to the projected spatial momentum, $P^\mu = \gamma^\mu_{~\alpha} p^\alpha$, via the relation $P^i = E V^i$. The worldline of the particle is parameterized by the coordinate time $t$ as $(t, X^i(t))$. Substituting these definitions into the 4-dimensional geodesic equation, we obtain the following set of evolution equations in the $3+1$ formalism as follows
\begin{eqnarray}
\frac{dX^i}{dt}&=&\alpha V^i-\beta^i,\label{geodesic_a}\\
\frac{dV^i}{dt}&=&\alpha \bigg(V^i(V^j\partial_j\ln \alpha - V^jK_{jk}V^k) + 2V^jK^i_{~j}\nonumber\\
               &-&  V^j~^3\Gamma^i_{jk}V^k\bigg)-\gamma^{ij}\partial_j\alpha - V^i\partial_j\beta^j.\label{geodesic_b}
\end{eqnarray}
And the evolution of the energy $E$ as follows
\begin{equation}
\frac{dE}{dt} = E(\alpha K_{jk}V^j V^k - V^j\partial_j \alpha).
\end{equation}
The coupled system of the dynamical metric \eqref{metric} and the geodesic equations \eqref{geodesic_a} and \eqref{geodesic_b} allows us to numerically integrate the geodesics and track the orbital evolution in the background of the scalarizing black hole.

The dynamical scalarization of a charged black hole is primarily governed by two key parameters: the initial black hole charge $Q$ and the coupling strength $\eta$. Previous studies have demonstrated that for a fixed charge, the system will relax to distinct scalarized black hole solutions depending on the value of $\eta$. While the scalar potential $V(\phi)$ is also known to significantly influence the scalarization properties \cite{Zhang:2022kbf,Zhang:2024wci}. In this work, we focus on the simplest case of a massive scalar field to isolate the effects of the coupling parameter.

In a previous study \cite{Zhang:2024wci}, we investigated the dynamical scalarization of charged black holes governed by a general axionic potential. Our analysis revealed that the scalar field mass term exerts the dominant influence on the scalarization dynamics, while higher-order interaction terms play a negligible role. Motivated by this finding, in the present work, we simplify the model by retaining only the mass term
\begin{equation}
V(\phi)=m^2\phi^2.
\end{equation}
In the numerical simulations presented below, we fix the initial black hole charge at $Q=0.9$. To investigate the dynamics of scalarization, we select the coupling parameter $\eta$ and the mass parameter as
\begin{equation}
\eta \phi_0=(0, 5, 10, 15, 20),
\end{equation}
\begin{equation}
m M=(0, 0.25, 0.50, 0.75, 1.00),
\end{equation}
where $\eta \phi_0$ is dimensionless and we set $\phi_0 = 1$.

It is important to note that during the numerical evolution, the spacetime coordinates undergo a dynamical transformation driven by the gauge conditions imposed on the lapse function and shift vector \cite{Alcubierre:2002kk}. Although the initial data is constructed in isotropic coordinates with a vanishing shift vector, the shift vector evolves dynamically to counteract the grid stretching associated with the collapsing slice. At late times, the coordinates settle into a quasi-stationary state \cite{Alcubierre:2011pkc}. We have explicitly verified that for a standard Reissner-Nordstr\"om black hole, the system transitions into this stationary gauge configuration within an evolution time of $t=100M_0$. Further details regarding this calibration can be found in Ref.~\cite{Zhang:2025lhm}.

\begin{figure*}[htbp]
		\includegraphics[width=0.8\linewidth]{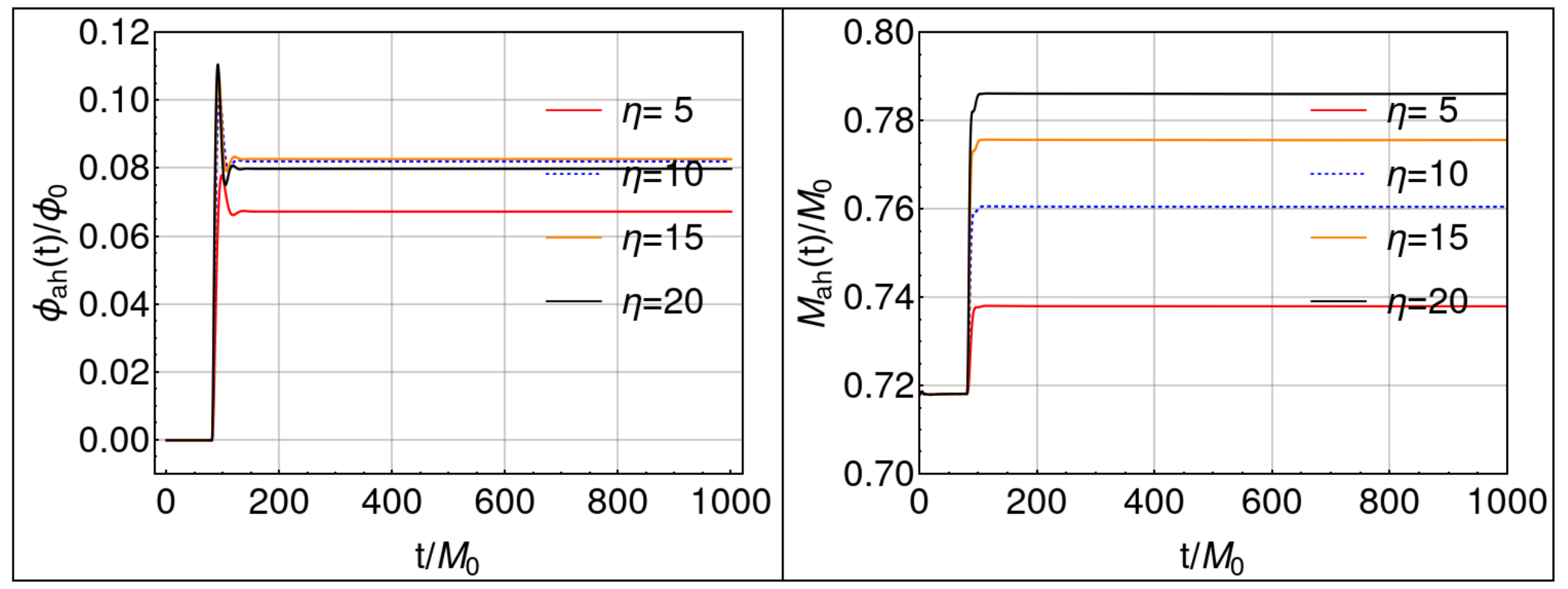}
	\caption{Time evolution of the value of scalar field on the apparent horizon $\phi_{ah}$ and the irreducible mass $M_{\text{ah}}$ of black hole, for different values of the coupling parameter $\eta$, with the scalar field mass parameter set to $m=0$.}
	\label{p_m_phi_ah}
    \end{figure*}

Once the spacetime coordinates have relaxed to a quasi-stationary state, the physical instability driven by the non-vanishing coupling $\eta$ takes over, causing the charged black hole to transition into a scalarized phase. We have previously detailed how this scalarization process modifies the irreducible mass and the scalar field value at the apparent horizon in Ref.~\cite{Zhang:2024wci}. It was shown that the final equilibrium state depends critically on the coupling strength $\eta$. For the sake of simplicity, in Fig.~\ref{p_m_phi_ah}, we restrict our presentation to the massless scalar field case ($m=0$) to illustrate these dynamics.

As a preliminary validation step, we first evolve a pure Reissner-Nordstr\"om black hole by setting the coupling parameter $\eta=0$ to verify the fidelity of our geodesic integration. In this limit, scalarization is suppressed, and the initially stable timelike circular orbits should remain unperturbed. However, as previously noted, the background spacetime still undergoes coordinate evolution due to gauge settling. To isolate physical orbital dynamics from these coordinate artifacts, we initialize the test particles in stable circular orbits only after the black hole has relaxed to a quasi-stationary state.

To comprehensively map the dynamical response of the spacetime geometry, we initialize a family of timelike circular geodesics. We select the timelike circular orbits with initial radii spanning the range $r/M_0 \in [r_\text{ISCO}+0.001, r_\text{max}]$, distributed at uniform intervals of $\Delta r/M_0=0.1$. As a control case, we first verify that for the non-scalarizing Reissner-Nordstr\"om black hole ($\eta=0$), these initially stable circular orbits maintain their circularity throughout the time evolution.

\begin{figure}[htbp]
		\includegraphics[width=\linewidth]{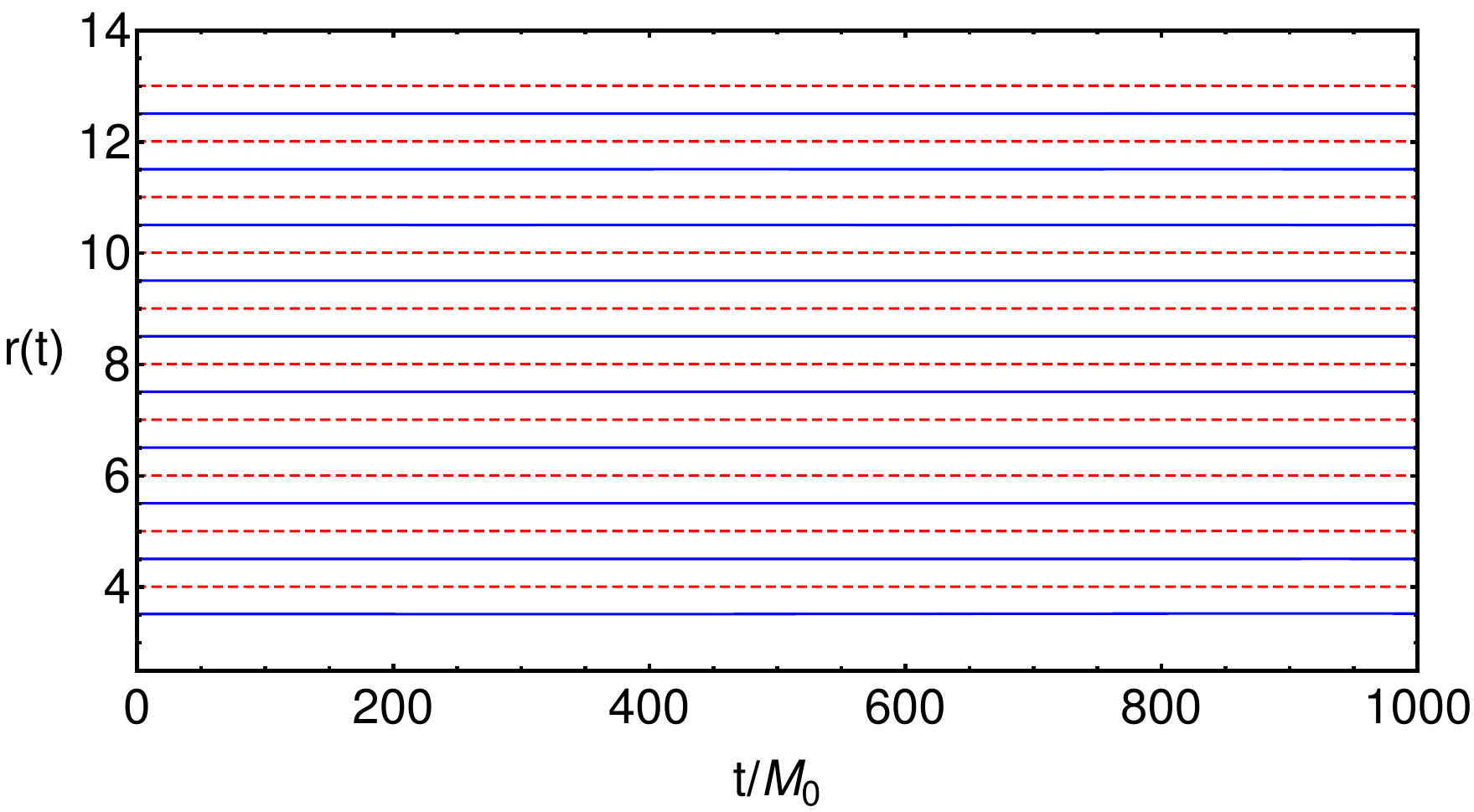}
	\caption{Time evolution of the radii for the initially circular timelike orbits in an evolving Reissner-Nordstr\"om black hole background with $\eta=0$.}
	\label{radii_evo_eta0}
\end{figure}

We compute the trajectories of these circular orbits alongside the evolution of the background spacetime. The resulting time dependence of the orbital radii is presented in Fig.~\ref{radii_evo_eta0}. As expected, our results confirm that for the non-scalarizing case, the initially stable circular orbits remain constant throughout the simulation. To assess the numerical precision of our code, we quantify the orbital error $O_\text{err}$ defined as the relative deviation of the evolved radius from its initial value
\begin{equation}
O_\text{err}=\frac{|r(t)-r(t_0)|}{r(t_0)}.
\end{equation}
We observe that for the innermost circular orbit considered in our sample, the maximum relative error $O_\text{err}$ reaches approximately $2\times 10^{-3}$. For orbits with larger radii, the numerical accuracy improves, with errors generally remaining on the order of $10^{-4}$ throughout the evolution.

Next, we investigate the impact of scalarization on initially stable timelike circular orbits within the Reissner-Nordström background. As previously discussed, the dynamic transition from a ``no-hair" to a ``hairy" state induces a time-dependent modification of the spacetime geometry. This geometric evolution fundamentally alters the effective potential, thereby changing the stability properties of the orbits established in the initial background. Furthermore, the specific characteristics of the final scalarized black hole are dictated by the choice of the coupling parameter $\eta$ and the scalar field mass $m$.

In Ref.~\cite{Zhang:2024wci}, we systematically investigated the evolution of the black hole irreducible mass and the scalar field profile during the dynamical scalarization process, considering a range of scalar masses and coupling strengths. Our analysis revealed that for a fixed coupling parameter $\eta$, increasing the scalar field mass $m$ leads to a suppression of the scalar field amplitude at the horizon of the final hairy black hole. This implies that the accumulation of scalar hair in the strong-field region is highly sensitive to the mass parameter, which in turn results in distinct background geometries for the final equilibrium state.

\begin{figure*}[!htbp]
		\includegraphics[width=\linewidth]{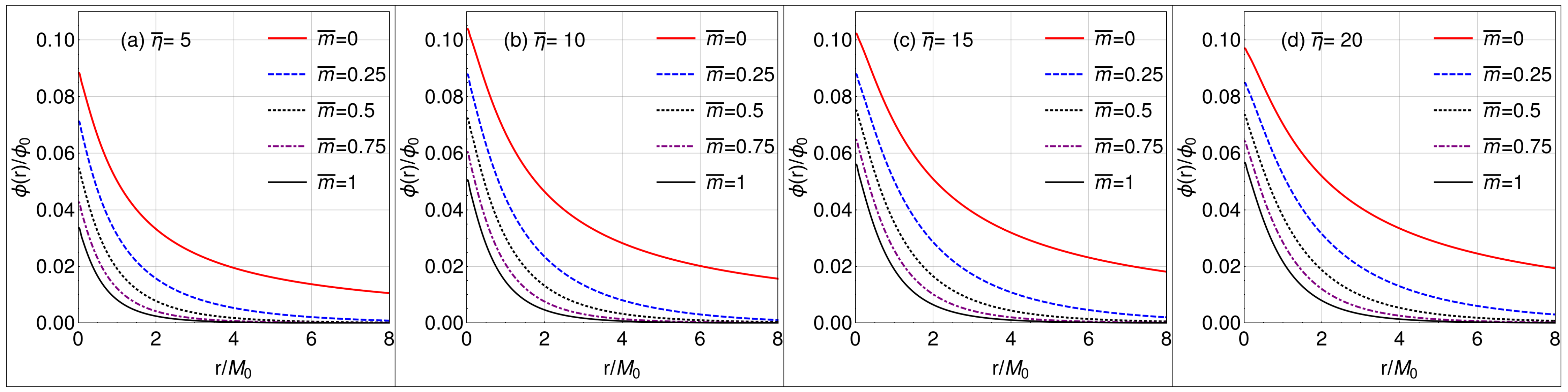}
	\caption{Radial distribution of the scalar field in the final scalarized black hole background ($t=500 M_0$).}
	\label{radialphi}
\end{figure*}

\begin{figure*}[!htbp]
		\includegraphics[width=\linewidth]{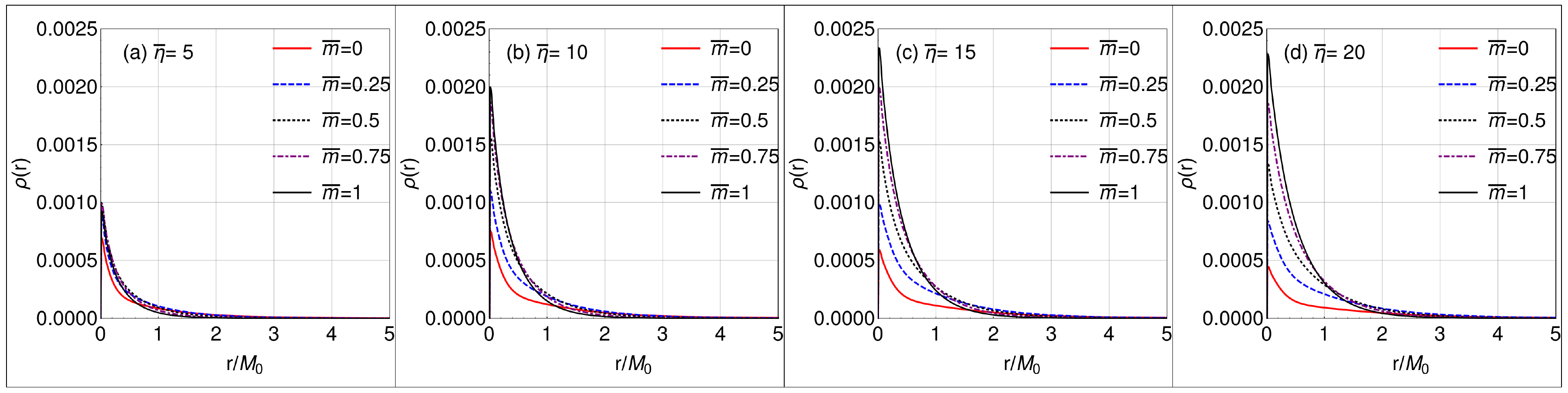}
	\caption{Radial distribution of the scalar field energy density in the final black hole background ($t=500 M_0$).}
	\label{radialrho}
\end{figure*}

To provide a physical basis for interpreting the orbital dynamics, we first analyze the spatial structure of the scalar field and its associated energy density in the final equilibrium state. Figures \ref{radialphi} and \ref{radialrho} display these radial profiles for various model parameters. We observe that the scalar field amplitude decays monotonically with increasing radius. This spatial fall-off implies a distance-dependent interaction: circular orbits with larger initial radii experience weaker perturbations, whereas those closer to the black hole are significantly more affected. Regarding the mass parameter $m$, we observe a distinct confinement effect.  While increasing $m$ generally suppresses the scalar field amplitude at large distances, it tends to concentrate the scalar field energy closer to the event horizon. In other words, the massive scalar hair becomes tightly confined to the strong-field region. This confinement mechanism suggests that, compared to the massless case, massive scalar fields will induce more drastic destabilization of orbits with small initial radii. As we demonstrate below, our full numerical evolutions strictly corroborate this physical intuition.

\begin{figure*}[!htbp]
		\includegraphics[width=\linewidth]{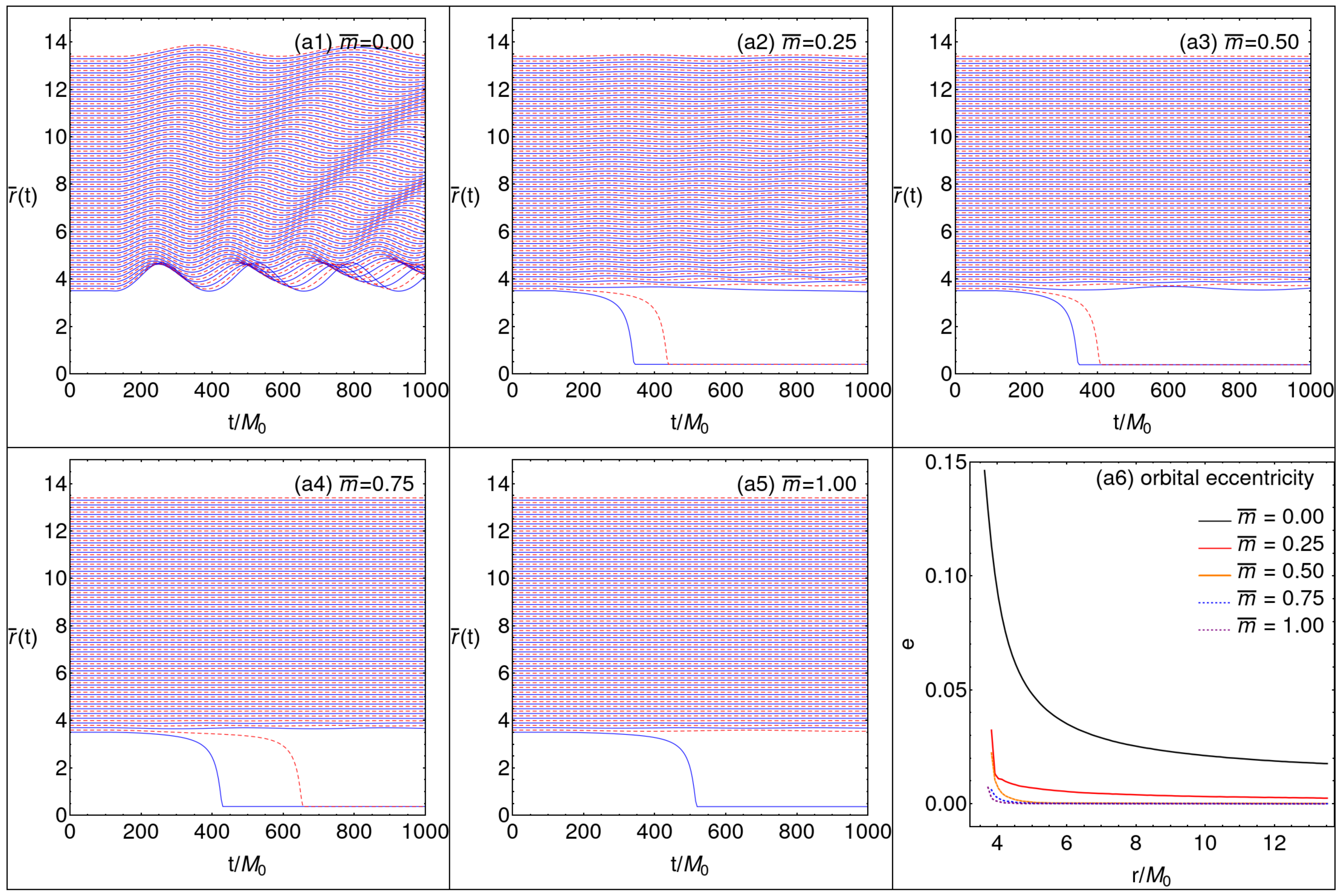}
	\caption{Time evolution of the orbital radii for initially stable circular orbits during scalarization with coupling parameter $\eta=5$ and scalar masses $m \in \{0, 0.25, 0.5, 0.75, 1.0\}$. The final orbital eccentricities induced by the scalarization are also displayed in subfigure (a6).}
	\label{orbit_evo_05}
\end{figure*}

\begin{figure*}[!htbp]
		\includegraphics[width=\linewidth]{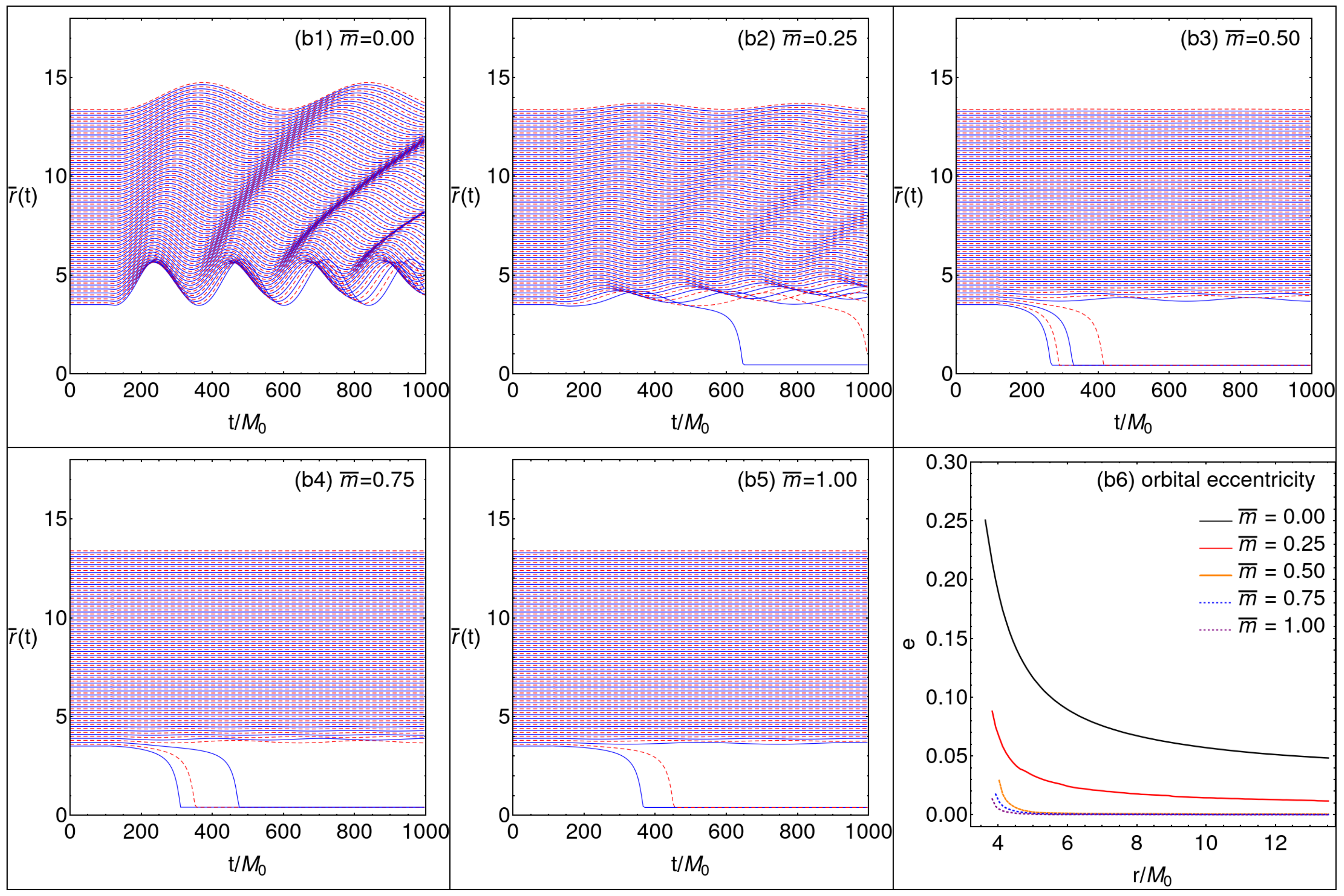}
	\caption{Time evolution of the orbital radii for initially stable circular orbits during scalarization with coupling parameter $\eta=10$ and scalar masses $m \in \{0, 0.25, 0.5, 0.75, 1.0\}$. The final orbital eccentricities induced by the scalarization are also displayed in subfigure (a6).}
	\label{orbit_evo_10}
\end{figure*}

\begin{figure*}[!htbp]
		\includegraphics[width=\linewidth]{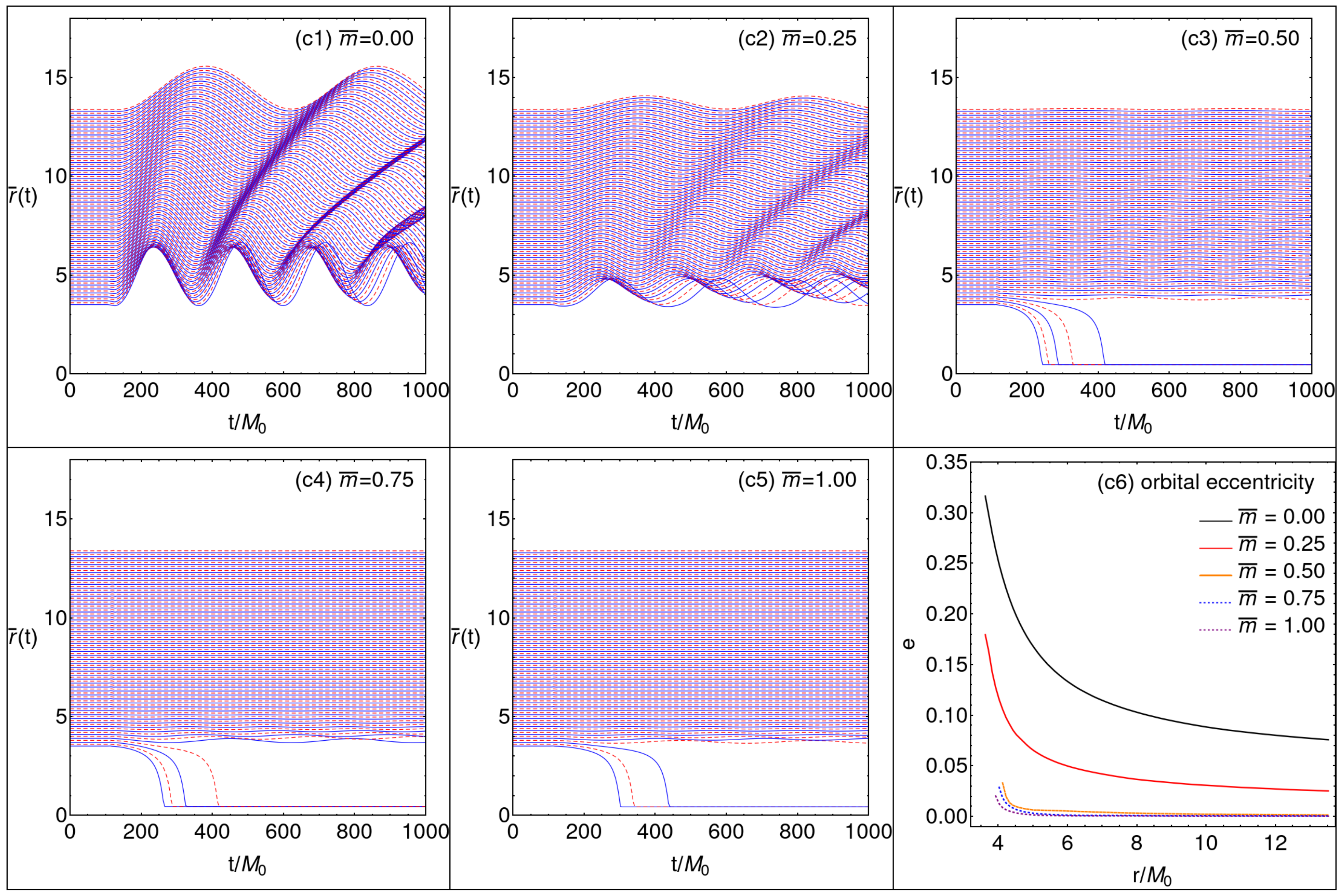}
	\caption{Time evolution of the orbital radii for initially stable circular orbits during scalarization with coupling parameter $\eta=15$ and scalar masses $m \in \{0, 0.25, 0.5, 0.75, 1.0\}$. The final orbital eccentricities induced by the scalarization are also displayed in subfigure (a6).}
	\label{orbit_evo_15}
\end{figure*}

\begin{figure*}[!htbp]
		\includegraphics[width=\linewidth]{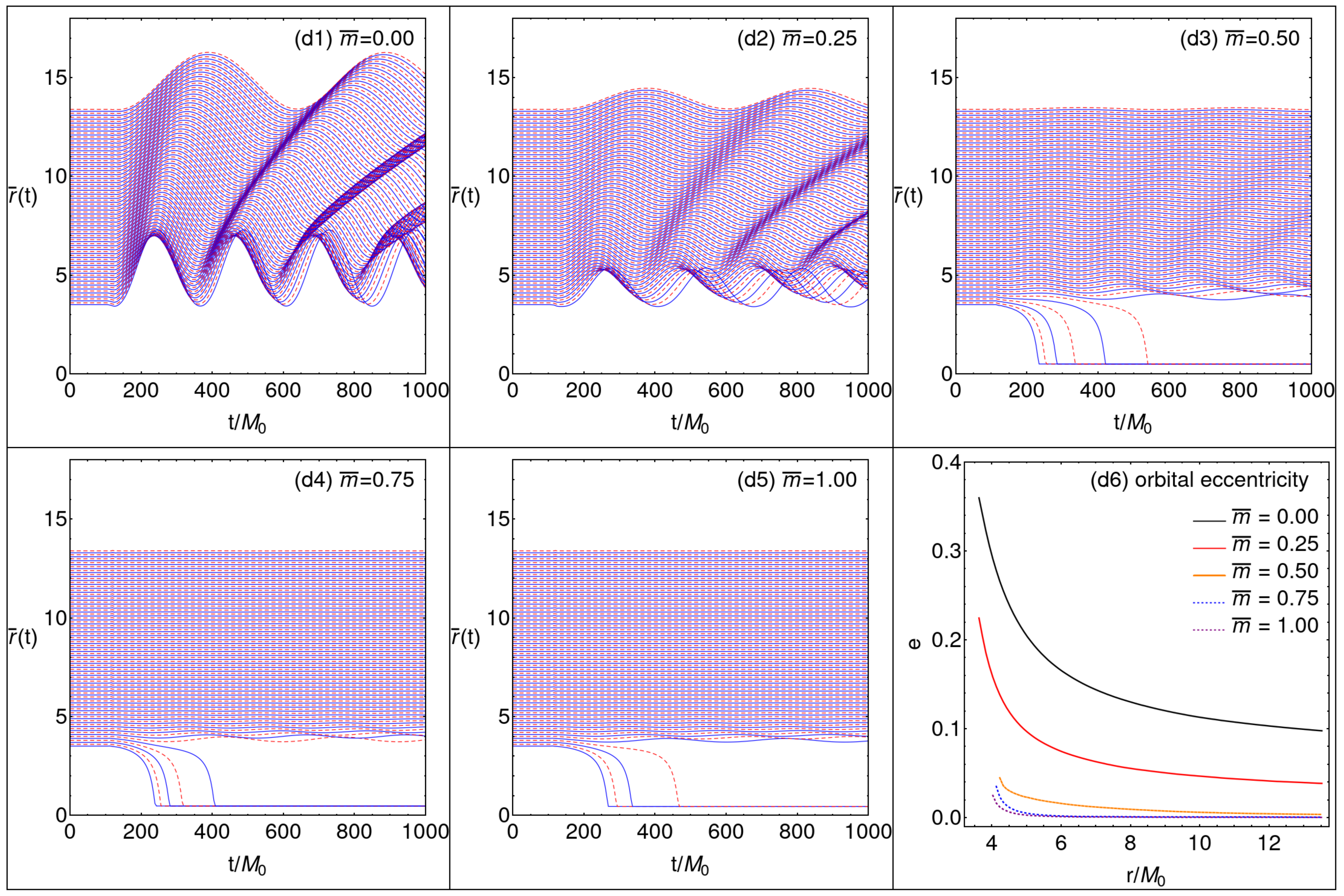}
	\caption{Time evolution of the orbital radii for initially stable circular orbits during scalarization with coupling parameter $\eta=20$ and scalar masses $m \in \{0, 0.25, 0.5, 0.75, 1.0\}$. The final orbital eccentricities induced by the scalarization are also displayed in subfigure (a6).}
	\label{orbit_evo_20}
\end{figure*}

Finally, we present the complete dynamical evolution of initially circular orbits during the black hole scalarization process. Figures \ref{orbit_evo_05}, \ref{orbit_evo_10}, \ref{orbit_evo_15}, and \ref{orbit_evo_20} display the results for varying scalar field masses $m$, each figure for a different value of the coupling strength $\eta$. We observe a distinct long-range effect in the massless limit: for a fixed $\eta$, the scalarization induces significant perturbations even on orbits with relatively large radii. Consequently, these trajectories lose their circularity and evolve into eccentric orbits around the final scalarized black hole. To quantify this deformation, the relationship between the induced eccentricity (measured in the final equilibrium background) and the initial orbital radius is explicitly plotted in panel (a6) of each figure.

Our analysis further reveals the critical role of the scalar mass parameter $m$. As $m$ increases, the scalarization exerts a significantly stronger destabilizing effect on inner orbits (those with small initial radii). This trend is corroborated by the relationship between the final eccentricity and the initial radius. Notably, in contrast to the massless case, massive scalar fields are more likely to force these inner orbits to plunge into the black hole. This behavior is consistent with the confinement of the scalar field energy density discussed earlier. Conversely, for outer orbits (large radii), the influence of the scalar field is suppressed as $m$ increases. This implies that for a sufficiently large mass parameter, the scalarization effects become localized near the event horizon, leaving distant orbits effectively unperturbed.

\section{Conclusions}\label{Conclusion}

In this work, we combined full numerical relativity simulations with the $3+1$ geodesic formalism to systematically investigate the interplay between scalar field dynamics and orbital motion during the scalarization of a charged black hole. Our analysis centered on the distinct roles played by two key parameters: the scalar field mass $m$ and the coupling strength $\eta$. By examining the radial profile of the scalar energy density and the resulting spacetime deformations, we have elucidated the mechanism by which the scalar mass regulates orbital evolution. Specifically, we demonstrated that the confinement of massive scalar fields leads to strong destabilization of inner orbits while suppressing effects at larger radii. These findings provide a robust theoretical framework for interpreting potential observational signatures of black hole scalarization in future astrophysical surveys.

To elucidate the physical origin of the orbital perturbations, we analyzed the spatial profile of the scalar field. For a fixed coupling $\eta$, increasing the scalar mass $m$ leads to a monotonic suppression of the global scalar field amplitude. However, the scalar energy density exhibits a contrasting behavior: it becomes increasingly concentrated near the event horizon. This behavior suggests that the scalar mass term confines the scalar field spatially, causing the scalar hair to become increasingly concentrated near the black hole. This distinction is pivotal for orbital dynamics: the massless scalar field ($m=0$) possesses a long-range profile that extends far from the black hole, thereby affecting distant orbits. In contrast, massive scalar fields are exponentially suppressed and localized within the strong-field region, limiting their dynamical influence to the immediate vicinity of the horizon.

In summary, our numerical results established a clear correlation between the scalar field properties (specifically the mass parameter) and the dynamical stability of the accretion flow. In the massless limit, the scalarization induces a long-range perturbation, driving even distant circular orbits into eccentric trajectories around the final equilibrium state. In contrast, the introduction of a scalar mass term fundamentally alters this behavior. As the mass parameter increases, the dynamical instability becomes increasingly confined to the strong-field region. Consequently, inner orbits (small radii) experience severe destabilization, often leading to a direct plunge into the horizon. Conversely, the influence on outer orbits is effectively suppressed; for sufficiently large scalar masses, the scalarization effects are localized near the event horizon, leaving distant orbits virtually unperturbed. This orbital response pattern directly mirrors the spatial confinement mechanism of the massive scalar field, confirming that the range of the scalar interaction dictates the scope of its astrophysical footprint.

\section{Acknowledgments}
	
This work was supported in part by the National Natural Science Foundation of China (Grants No. 12575055, 12475056, 12475055, and 12247101), the Fundamental Research Funds for the Central Universities (Grant No. lzujbky-2025-jdzx07), the Natural Science Foundation of Gansu Province (No. 22JR5RA389, No.25JRRA799), and the `111 Center' under Grant No. B20063, the Gansu Province's Top Leading Talent Support Plane. Yu-Peng Zhang was supported by ``Talent Scientific Fund of Lanzhou University".

\end{document}